\input harvmac
\noblackbox
 
\font\ticp=cmcsc10
 
\def\Title#1#2{\rightline{#1}\ifx\answ\bigans\nopagenumbers\pageno0\vskip1in
\else\pageno1\vskip.8in\fi \centerline{\titlefont #2}\vskip .5in}

\font\ticp=cmcsc10
\font\ttsmall=cmtt10 at 8pt

\input epsf
\ifx\epsfbox\UnDeFiNeD\message{(NO epsf.tex, FIGURES WILL BE
IGNORED)}
\def\figin#1{\vskip2in}
\else\message{(FIGURES WILL BE INCLUDED)}\def\figin#1{#1}\fi
\def\ifig#1#2#3{\xdef#1{fig.~\the\figno}
\goodbreak\topinsert\figin{\centerline{#3}}%
\smallskip\centerline{\vbox{\baselineskip12pt
\advance\hsize by -1truein\noindent{\bf Fig.~\the\figno:} #2}}
\bigskip\endinsert\global\advance\figno by1}

%
%
\def\[{\left [}
\def\]{\right ]}
\def\({\left (}
\def\){\right )}
\def\p{\partial}
\def\R{{\bf R}}
\def\S{{\bf S}}

\def\om{\omega}
\def\a{\alpha}

\def\lam{\lambda}

\def\dta{\( {\p \over \p t} \)^a}
\def\dya{\( {\p \over \p y} \)^a}
\def\dza{\( {\p \over \p z} \)^a}

\def\dVa{\( {\p \over \p V} \)^a}
\def\dphia{\( {\p \over \p \phi} \)^a}
\def\dlama{\( { d \over d \lam} \)^a}
\def\dlamb{\( { d \over d \lam} \)^b}

\def\ie{{\it i.e.}}
\def\godel{G{\" o}del}
\def\clg{G{\" o}del}
\def\clp{pp-wave}
\def\t{\tau}
\def\ku{\xi_{e_u}}
\def\kv{\xi_{e_v}}
\def\ki{\xi_{e_i}}
\def\kj{\xi_{e_j}}
\def\kk{\xi_{e_k}}
\def\kone{\xi_{e_1}}
\def\kis{\xi_{{e_i}^*}}
\def\kjs{\xi_{{e_j}^*}}
\def\kks{\xi_{{e_k}^*}}
\def\kones{\xi_{{e_1}^*}}
\def\ktwos{\xi_{{e_2}^*}}
\def\krotij{\xi_{M_{ij}}}
\def\krotkl{\xi_{M_{kl}}}
\def\krotil{\xi_{M_{il}}}
\def\krotjl{\xi_{M_{jl}}}
\def\krotik{\xi_{M_{ik}}}
\def\krotjk{\xi_{M_{jk}}}

\def\cu{\cos(\mu \, u)}
\def\cus{\cos^2(\mu \, u)}

\def\su{\sin(\mu \, u)}

\def\sus{\sin^2(\mu \, u)}


\lref\bmn{
D.~Berenstein, J.~M.~Maldacena and H.~Nastase,
{\it Strings in flat space and pp waves from N = 4 super Yang Mills},
JHEP {\bf 0204}, 013 (2002)
[arXiv:hep-th/0202021].}

\lref\bfhp{
M.~Blau, J.~Figueroa-O'Farrill, C.~Hull and G.~Papadopoulos,
{\it A new maximally supersymmetric background of IIB superstring theory},
JHEP {\bf 0201}, 047 (2002)
[arXiv:hep-th/0110242].}

\lref\host{
G.~T.~Horowitz and A.~R.~Steif,
{\it Space-Time Singularities In String Theory},
Phys.\ Rev.\ Lett.\  {\bf 64}, 260 (1990).}

\lref\amkl{
D.~Amati and C.~Klimcik,
{\it Nonperturbative Computation Of The Weyl Anomaly For A Class Of 
Nontrivial Backgrounds},
Phys.\ Lett.\ B {\bf 219}, 443 (1989).}

\lref\maro{
D.~Marolf and S.~F.~Ross,
{\it Plane waves: To infinity and beyond!},
Class.\ Quant.\ Grav.\  {\bf 19}, 6289 (2002)
[arXiv:hep-th/0208197].}

\lref\penrose{
R.~Penrose,
{\it Any spacetime has a planewave as a limit}, in
 Differential geometry and relativity, pp 271-275,
Reidel, Dordrecht, 1976.}

\lref\penr{
R.~Penrose,
{\it A Remarkable Property Of Plane Waves In General Relativity},
Rev.\ Mod.\ Phys.\  {\bf 37}, 215 (1965).}

\lref\bena{
D.~Berenstein and H.~Nastase,
{\it On lightcone string field theory from super Yang-Mills and holography},
[arXiv:hep-th/0205048].}

\lref\malmaoz{
J.~Maldacena and L.~Maoz,
{\it Strings on pp-waves and massive two dimensional field theories},
[arXiv:hep-th/0207284].}

\lref\bghv{
E.~K.~Boyda, S.~Ganguli, P.~Ho\v{r}ava and U.~Varadarajan,
{\it Holographic protection of chronology in universes of the \godel\ type},
Phys.\ Rev.\ D {\bf 67}, 106003 (2003)
[arXiv:hep-th/0212087].
}

\lref\hata{T.~Harmark and T.~Takayanagi,
{\it Supersymmetric \godel\ universes in string theory},
[arXiv:hep-th/0301206].
}

\lref\figsim{J.~Figueroa-O'Farrill and J.~Simon,
{\it Generalized supersymmetric fluxbranes},
JHEP {\bf 0112}, 011 (2001)
[arXiv:hep-th/0110170].
}

\lref\geho{R.~Geroch and G.~T.~Horowitz,
{\it Global structure of spacetimes}, in General Relativity: An 
Einstein Centenary Survey, eds. S~.W.~Hawking and W.~Israel, 
Cambridge Univ Press.
}

\lref\michelson{J.~Michelson,
{\it (Twisted) toroidal compactification of pp-waves},
Phys.\ Rev.\ D {\bf 66}, 066002 (2002)
[arXiv:hep-th/0203140].
}

\lref\justin{J.~R.~David,
{\it Plane waves with weak singularities},
[arXiv:hep-th/0303013].
}

\lref\simon{
J.~Simon,
{\it Null orbifolds in AdS, time dependence and holography},
JHEP {\bf 0210}, 036 (2002)
[arXiv:hep-th/0208165].
}

\lref\hael{S.~W.~Hawking and G.~F.~R.~Ellis, The large scale structure
of space-time, Cambridge Univ Press.}

\lref\bdpr{
D.~Brecher, P.~A.~DeBoer, D.~C.~Page and M.~Rozali,
{\it Closed timelike curves and holography in compact plane waves},
[arXiv:hep-th/0306190].
}

\lref\bero{
I.~Bena and R.~Roiban,
{\it Supergravity pp-wave solutions with 28 and 24 supercharges},
Phys.\ Rev.\ D {\bf 67}, 125014 (2003)
[arXiv:hep-th/0206195].
}

\lref\michelsonb{
J.~Michelson,
{\it A pp-wave with 26 supercharges},
Class.\ Quant.\ Grav.\  {\bf 19}, 5935 (2002)
[arXiv:hep-th/0206204].
}

\lref\flsa{
J.~L.~Flores and M.~Sanchez,
{\it Causality and conjugate points in general plane waves},
Class.\ Quant.\ Grav.\  {\bf 20}, 2275 (2003)
[arXiv:gr-qc/0211086].
}

\lref\gauntlett{
J.~P.~Gauntlett, J.~B.~Gutowski, C.~M.~Hull, S.~Pakis and H.~S.~Reall,
{\it All supersymmetric solutions of minimal supergravity in 
five  dimensions},
[arXiv:hep-th/0209114].
}

\lref\cstruct{
V.~E.~Hubeny and M.~Rangamani,
{\it Causal structures of pp-waves,}
JHEP {\bf 0212}, 043 (2002)
[arXiv:hep-th/0211195].
}

\lref\causc{
D.~Marolf and S.~F.~Ross,
{\it A new recipe for causal completions},
[arXiv:gr-qc/0303025].
}

\lref\elsch{
G.~F.~Ellis and B.~Schmidt,
{\it Singular space-times},
Gen. Rel. Gravit. {\bf 8}, 915 (1977).
}

\lref\tip{
F.~Tipler,
{\it Singularities in conformally flat spacetimes},
Phys. Lett. {\bf A64}, 8 (1977).
}

\lref\fapap{
J.~Figueroa-O'Farrill and G.~Papadopoulos,
{\it Homogeneous fluxes, branes and a maximally supersymmetric
solution of  M-theory},
JHEP {\bf 0108}, 036 (2001)
[arXiv:hep-th/0105308].
}

\lref\ruts{
J.~G.~Russo and A.~A.~Tseytlin,
{\it A class of exact pp-wave string models with interacting light-cone 
gauge actions},
JHEP {\bf 0209}, 035 (2002)
[arXiv:hep-th/0208114].
}

\lref\baso{
I.~Bakas and J.~Sonnenschein,
{\it On integrable models from pp-wave string backgrounds},
JHEP {\bf 0212}, 049 (2002)
[arXiv:hep-th/0211257].
}
\lref\btz{
M.~Banados, C.~Teitelboim and J.~Zanelli,
{\it The Black Hole In Three-Dimensional Space-Time},
Phys.\ Rev.\ Lett.\  {\bf 69}, 1849 (1992)
[arXiv:hep-th/9204099];
M.~Banados, M.~Henneaux, C.~Teitelboim and J.~Zanelli,
{\it Geometry of the (2+1) black hole},
Phys.\ Rev.\ D {\bf 48}, 1506 (1993)
[arXiv:gr-qc/9302012].
}

\lref\herdeiro{
C.~A.~Herdeiro,
{\it Spinning deformations of the D1-D5 system and a geometric resolution 
of  closed timelike curves},
[arXiv:hep-th/0212002].
}

\lref\herdeiroold{
C.~A.~Herdeiro,
{\it Special properties of five dimensional BPS rotating black holes},
Nucl.\ Phys.\ B {\bf 582}, 363 (2000)
[arXiv:hep-th/0003063].
}


%
\baselineskip 16pt
\Title{\vbox{\baselineskip12pt
\line{\hfil SU-ITP-03/20}
\line{\hfil UCB-PTH-03/18}
\line{\hfil LBNL-53482}
\line{\hfil DCPT-03/39}
\line{\hfil \tt hep-th/0307257} }}
{\vbox{
{\centerline{}  Causal inheritance in plane wave quotients
}}}

\centerline{\ticp Veronika E. Hubeny$^a$,
 Mukund Rangamani$^{b,c}$  and Simon F. Ross$^{d}$\footnote{}{\ttsmall
veronika@itp.stanford.edu,  
mukund@socrates.berkeley.edu,  
S.F.Ross@durham.ac.uk}}

\bigskip

\centerline {\it $^a$
Department of Physics, Stanford University, Stanford, CA 94305, USA} 
\centerline{\it $^b$ Department of Physics, University of California,
Berkeley, CA 94720, USA} 
\centerline{\it $^c$ Theoretical Physics Group, LBNL, Berkeley, CA 94720, USA}
\centerline{\it $^d$ Department of Mathematical Sciences, University 
of Durham, Durham, DH1 3LE UK}


\bigskip
\centerline{\bf Abstract}
\bigskip

\noindent

We investigate the appearance of closed timelike curves in quotients
of plane waves along spacelike isometries. First we formulate a
necessary and sufficient condition for a quotient of a general
spacetime to preserve stable causality. We explicitly show that the
plane waves are stably causal; in passing, we observe that some
pp-waves are not even distinguishing. We then consider the
classification of all quotients of the maximally supersymmetric
ten-dimensional plane wave under a spacelike isometry, and show that
the quotient will lead to closed timelike curves iff the isometry
involves a translation along the $u$ direction. The appearance of
these closed timelike curves is thus connected to the special
properties of the light cones in plane wave spacetimes. We show that
all other quotients preserve stable causality.

\Date{July, 2003}
%


\newsec{Introduction}


Plane waves are a class of metrics of great interest for string
theory, for a number of reasons: Certain plane waves provide maximally
supersymmetric backgrounds for string propagation \bfhp. Furthermore,
plane waves can be obtained as the neighbourhood of a null geodesic in
a generic spacetime via the Penrose limit \penrose, which allows us
to use plane waves to explore and test the dualities between string
theory in certain backgrounds and supersymmetric gauge theories
\bmn. These spacetimes also have vanishing curvature invariants, which
makes them exact backgrounds for string propagation to all orders in
$l_s$ \amkl, \host.

It has recently been noted that the maximally supersymmetric plane
wave of \bfhp\ is T-dual \bghv, \hata\
to a supersymmetric G\"odel-like universe \gauntlett. 
This might seem surprising; the plane wave is a stably causal
spacetime, and is hence free of causal pathologies; the G\"odel-like
solution, on the other hand, has closed timelike curves (CTCs) passing
through every point of the spacetime. We are used to the idea that
T-duality can change the geometrical properties of the solution
considerably; but is it really possible for T-duality to relate a
causally well-behaved spacetime to one with CTCs\foot{If true, this
would necessarily imply that string theory can be defined consistently
on the background with CTCs, since the two geometries are given by the
same worldsheet CFT.  This would seem to rule out the possibility that
some general mechanism in string theory forbids the existence of
CTCs.}?

In fact, the CTCs are not introduced by T-duality, but rather by
quotienting. The plane wave does not have any compact circle
directions. We must first quotient the spacetime by some suitable
spacelike isometry to obtain a geometry with a finite-radius circle,
which we can then consider the T-duality along. 
As we will discuss in section 2, it is this quotient that introduces
the CTCs; they are present on both sides of the T-duality. That
spacelike quotients can produce CTCs was first observed in \herdeiroold; 
for the specific case of \godel, this was previously
discussed in \herdeiro, \bghv, \bdpr.

The true surprise, then, is that the quotient of a causally
well-behaved spacetime along an everywhere spacelike isometry can lead
to a spacetime with closed timelike curves. The main purpose of the
present paper is to explore this discovery in more detail. We will
follow two lines of development: we want to understand in general when
such violations of causality conditions can occur, and we want to see
why such a breakdown occurred for the plane wave in particular.

Firstly, we will consider the general question of quotients and
causality. We will review the hierarchy of causality conditions that a
spacetime may satisfy, and explore the conditions under which they are
inherited by a quotient of the spacetime by some subgroup of the
isometry group (we focus on the case of a one-parameter subgroup,
\ie, quotients generated by a single isometry). It is clear that for
any causality condition to be inherited, it is necessary that the
isometry we quotient along be everywhere spacelike. However, this is
not a sufficient condition. Our main result in this general part of
the discussion is to give a new necessary and sufficient condition
under which a quotient of a stably causal spacetime inherits the
property of stable causality\foot{Stable causality is the strongest
of the usual causality conditions. It is equivalent to the existence of a
globally-defined time function on the spacetime.}.

To make these methods applicable to the case at hand, we will
investigate the causality properties of plane waves and pp-waves. We
will give an explicit demonstration that all plane waves are stably
causal; this appears to be a known result \geho, but we are unaware of
any previous explicit demonstration.  We will also discuss the
extension to pp-waves, following \flsa. We will see that they divide
into two classes: either they are stably causal, or they are causal
but fail to be distinguishing\foot{That is, not all distinct points have
distinct pasts and futures.}.

We then consider the particular question of which quotients of plane
waves preserve causality conditions, and why. We will consider the
classification of the spacelike isometries of the maximally symmetric
plane wave geometry in ten dimensions \bfhp\ (henceforth BFHP plane
wave). This is a problem of interest in its own right: the
classification of spacelike isometries of flat space led to
interesting discoveries \figsim, and there have also been related
investigations of AdS \simon. In the case of plane waves, several
authors have investigated possible quotients of the BFHP plane wave
\michelson, \bghv, \hata, but a unified treatment has so
far been lacking. We will show that the spacelike isometries of the
BFHP plane wave fall into two classes: the \clg\ class which leads to
quotient spacetimes with CTCs, and the \clp\ class which leads to
stably causal spacetimes; on Kaluza-Klein reduction to nine
dimensions, quotients in the \clp\ class become pp-waves.
The fact that the quotients in the \clg\ class lead to CTCs turns out
to be simply related to the special properties of the light cones in
the plane wave spacetime.

\subsec{Quotients with CTCs}

To motivate the ensuing general discussion, we will begin by
exhibiting two concrete examples wherein upon performing the quotient we
generate CTCs (see also related examples in \herdeiroold, \bdpr).  The
prototypical example involves a plane wave; for simplicity, we
consider here a 4-dimensional maximally symmetric plane wave, which
for convenience we write in spherical polar coordinates,
\eqn\fourbfhp{
ds^2 = -2 \, du \, dv - \mu^2 r^2 \, du^2 + dr^2 + r^2 \, d\theta^2 \ .
}
To exhibit the isometry explicitly, consider the change of coordinates
\eqn\coordcha{
   u = t + y,                \qquad 
   \theta = \phi - \mu (t + y),     \qquad
   v = {1 \over 2} \, (t - y) \ ,
}
which casts \fourbfhp\ in the form
\eqn\fourbipolar{
ds^2 = - dt^2 + dy^2 + dr^2 + r^2 \, d\phi^2 - 2 \, \mu \, r^2 \, d\phi \, 
(dt + dy) \ .
}
Note that $\phi$ is a periodic coordinate: $\phi \sim \phi +
2\pi$. Now consider a quotient of this spacetime along the orbits of
the Killing vector $\dya$, identifying $y \sim y + 2\pi R$ to
form a compact circle of some radius $R$. Since $\dya$ is a
spacelike vector, the `obvious' closed curve is not a CTC. However,
since $\phi$ is periodic, the curve $(t,y,r,\phi) = (t_0,2\pi n R \, 
\lambda,r_0,2\pi \lambda)$, where $\lambda \in [0,1]$, is also closed
for every integer $n$ (this curve only closes after winding $n$ times
around the compact $y$ direction). The tangent to this curve,
\eqn\ctctan{ \dlama  = 2 \pi n R \, \dya +
2\pi \, \dphia \ , }
has norm 
\eqn\ctcnorm{ g_{a b} \, \dlama \, \dlamb = 4\pi^2
(n^2 R^2 + r_0^2 - 2 \mu n R r_0^2).
}
Thus, if we choose $n$ such that $2 \mu n R >1$, then $\dlama$ 
is timelike for sufficiently large $r_0$, and hence the
quotient spacetime contains CTCs. Thus in this prototypical example it
is clear that quotienting a stably causal spacetime by an everywhere
spacelike isometry can lead to CTCs. Requiring that the isometry be
everywhere spacelike is necessary to avoid CTCs, but it is not always
sufficient. 

Another simple example involves the three-dimensional
Banados-Teitelboim-Zanelli (BTZ) black hole
\btz. Here the metric is 
\eqn\btzmet{ ds^2 = - (r^2 - r_+^2)\,  dt^2 + {dr^2 \over (r^2 - r_+^2)}
+ r^2 \, d\phi^2 \ . }
If we consider the Killing vector $\xi^a = \a \, \dta + \dphia$,
it will have norm $\xi^a \, \xi_a = -\a^2 \, (r^2-r_+^2) + r^2$, so it is
everywhere spacelike in the BTZ spacetime if $\a^2 < 1$. Consider a
quotient along this Killing vector, identifying $(t,r,\phi) \sim
(t+2\pi \a R, r, \phi + 2\pi R)$ for some $R$. Again, since $\phi$ is
periodic, $\phi \sim \phi + 2\pi$, this will also produce
identifications of the form $(t,r,\phi) \sim (t+2\pi n \a R, r, \phi +
2\pi (nR - m))$, for $n,m$ integers. This produces CTCs in the region
$r>r_+$, as we can choose $n$ and $m$ such that $ -\a^2 (r^2- r_+^2) +
r^2 [1 - m/(nR)]^2 < 0$ \foot{Note that these CTCs are unrelated to
the CTCs which occur if we include the region $r<0$ in the BTZ metric;
the further quotient we consider here produces CTCs which lie entirely
in the `exterior' region $r>r_+$.}.

This example differs from the previous one in two interesting
ways. Firstly, in the plane wave case, it was observed in \bdpr\ that
the identifications produce CTCs but not closed timelike
geodesics. This is no longer the case in our BTZ example; it is easy
to find a timelike geodesic which connects the two points $(t,r,\phi)$
and $(t+2\pi n \a R, r, \phi + 2\pi (nR - m))$, by travelling out to
larger $r$ and reflecting off the radial potential in BTZ. Thus, we
see that there is no obstruction in principle to obtaining closed
timelike geodesics from such quotients. Secondly, the appearance of
CTCs in the plane wave is, as we shall see later, intimately connected
to the special properties of the light cone in the plane wave
spacetime. In the BTZ spacetime, on the other hand, there is nothing
unusual about the light cones in the region $r > r_+$, and one can
easily construct similar examples where a spacelike quotient of a
globally hyperbolic spacetime produces CTCs. Thus, this phenomenon
seems less dependent on the special features of the causal structure
of plane waves.

However, the appearance of CTCs in the BTZ geometry
also depends on a special feature of this metric: that BTZ is itself a
quotient. One can think of this example as the quotient of AdS$_3$ by
a group generated by two Killing vectors, $\dphia$ and $\xi^a = \a \,
\dta + \dphia$. Since they differ by a timelike vector,
it is perhaps not surprising that quotienting by both actions gives
rise to CTCs. Put another way, the key feature of the BTZ example is
that it has a spacelike isometry with non-contractible $\S^1$
orbits. There are clearly many other examples of spacetimes with such
$\S^1$s which will produce CTCs under an appropriate spacelike
quotient; however, these seem somewhat artificial examples, and are
less interesting than the plane wave case, where the orbits of
$\dphia$ are contractible, and the appearance of CTCs comes from the
causal structure.

\newsec{Causal inheritance in general}

An obvious question prompted by the above examples is, {\it does there
exist} a sufficient condition for a given quotient of some arbitrary
spacetime to be free of CTCs? In this section, we will see that there
is in fact a simple and elegant condition for stable causality to be
inherited under a quotient, which is both necessary and
sufficient. Since stable causality is a stronger requirement than the
absence of CTCs, as an immediate consequence this provides the desired
sufficient condition for the absence of CTCs (albeit only in the case
where the `parent' spacetime is stably causal).

\subsec{Causality conditions}

To make the discussion comprehensible to readers unfamiliar with the
intricacies of causality, we will briefly review the canonical
hierarchy of causality conditions. This is a `pocket guide'; for a
more extensive discussion of these conditions, see, e.g., \hael. These
conditions are ordered by increasing strength, so each condition
implies all the previous ones. 

A spacetime that is free of closed timelike curves is said to satisfy
the {\it chronology} condition. Furthermore, a spacetime is said to be
{\it causal} if it contains no closed non-spacelike curves. The causal
condition is the weakest condition that it seems desirable to impose
on spacetimes. However, metrics satisfying this condition can still
contain `almost closed' timelike curves, which may still lead to
pathologies. (What we mean formally by this is that an infinitesimal
perturbation of the metric may be sufficient to produce CTCs.) It is
therefore useful to consider stronger conditions on the causal
structure.

We call a spacetime $M$ {\it weakly distinguishing} if for $p,q \in
M$, $p = q$ whenever $I^+(p) = I^+(q)$ and $I^-(p) = I^-(q)$. (Here
$I^+(p)$ is the chronological future of the point $p$ --- the set of
points reachable by future-directed timelike curves from $p$ --- and
$I^-(p)$ is the chronological past of $p$). Similarly, the spacetime is {\it
future-distinguishing} if $p = q$ whenever $I^+(p) = I^+(q)$, {\it
past-distinguishing} if $p = q$ whenever $I^-(p) = I^-(q)$, and simply
{\it distinguishing} if $p = q$ whenever either $I^+(p) = I^+(q)$ or
$I^-(p) = I^-(q)$. Clearly a spacetime satisfying any of these
conditions cannot contain CTCs, since any two points $p$ and $q$ on a
CTC $\gamma$ will have $I^+(p) = I^+(q)$ and $I^-(p) = I^-(q)$. 

A spacetime is called {\it strongly causal} if for every point $p$ in
the spacetime manifold, and every neighbourhood $O$ of $p$, there
exists a neighbourhood $V$ of $p$ contained in $O$ which no causal
curve intersects more than once. Basically, we require that causal
curves passing sufficiently near $p$ do not come arbitrarily close to
being closed curves. The strong causality condition is necessary for
it to be possible to define the causal boundary of a spacetime using
the technique of \causc. It also implies that the Alexandrov topology
of the spacetime (the topology determined by the causal structure)
agrees with the manifold topology.

A spacetime is called {\it stably causal} if the spacetime metric
$g_{\mu \nu}$ has an open neighbourhood in the family of continuous
metrics such that none of the metrics in the neighbourhood of the
given metric admit closed timelike curves.  Said differently,
perturbing the spacetime metric by opening out light cones at every
point should not lead to closed timelike curves.  In a theory of
quantum gravity, where the metric is subject to quantum fluctuation,
the stable causality condition is likely to be the minimum required to
avoid possible pathologies associated with CTCs. It is therefore
usually assumed that only stably causal spacetimes are of physical
relevance; this includes most of the usual standard backgrounds, such
as flat space, anti-de Sitter space, plane waves (as we will see below), and
p-branes. In fact, it is highly unusual to encounter a spacetime which
is causal but not stably causal in practice (clearly such spacetimes
are by definition a subset of measure zero in the space of continuous
metrics). One of the interesting features of our discussion of the
pp-waves is that they provide the first really natural example of such
causally intermediate spacetimes.

Stable causality is equivalent to a much simpler requirement: it can
be proved \hael\ that a spacetime is stably causal {\it iff } there is
a smooth function $\t$ on the spacetime manifold whose gradient
$\nabla_a \, \t$ is everywhere timelike.  Such a function $\t$ is
called a {\it time function}, and it behaves as a good measure of time
on the entire spacetime manifold, in the sense that it increases along
every future directed causal curve. A time function also determines a
natural foliation of the spacetime by spacelike surfaces, which are
surfaces of simultaneity with respect to the time function. The
definition of the time function clearly contains a lot of freedom;
given a time function, we can construct another by adding to it a
sufficiently small multiple of any smooth function on the spacetime.
Much of our subsequent discussion will be concerned with the
definition of time functions on stably causal spacetime, and on the
plane waves in particular.

If the spacetime additionally has a timelike Killing vector field
$k^a$, one can specify a class of time functions by 
requiring that the Lie                              
derivative of $\t$ along $k^a$ 
$\CL_k \t = 1$, although this is still far from determining $\t$
uniquely. For familiar examples, such as flat space, Anti de Sitter
spacetimes, and the Einstein Static Universe (ESU), this condition is
satisfied, for example, by the canonical time function $\t = t$ in
terms of which the timelike Killing vector is $\dta$. However, the
existence of a timelike Killing vector and stable causality are
unrelated. For example, the compactification of flat space along the
timelike isometry $\dta$ preserves the Killing vector, but clearly
violates stable causality. The point is that $\t = t$ is not a smooth
function on the quotient; it must have a discontinuity somewhere.

While there is no constructive procedure for determining a time
function for a given spacetime, there are various strategies at our
disposal to find a candidate time function. If the given spacetime can
be conformally embedded in a spacetime for which we know a time
function, such as the ESU, then the same function must be a time
function for the spacetime in question. This simply stems from the
fact that causal properties of a spacetime are conformally
invariant. Similarly, if we can find a fiducial metric $g_{\mu
\nu}^{(fid)}$ such that all timelike curves in our physical metric are
also timelike curves in the fiducial metric (so that the light cones
of the fiducial metric `lie outside' those of the physical metric),
then any time function for the fiducial metric is also a time function
for the physical metric. This is clear from the definition of a stably
causal spacetime.

 Finally, there is one stronger causality condition, which we include
for completeness: a spacetime is {\it globally hyperbolic} if it has a
global Cauchy surface. That is, there must be some spacelike surface
$\Sigma$ such that for any point $p$ in the manifold, every causal
curve through $p$ meets $\Sigma$ exactly once. This condition implies
stable causality. To see this, note that we can define a time function
$\t$ by setting $\t(p)$ equal to the volume of the intersection of
$I^-(p)$ with $\Sigma$ minus the volume of the intersection of
$I^+(p)$ with $\Sigma$. While global hyperbolicity simplifies various
constructions on the spacetime, imposing global hyperbolicity would be
too restrictive: important examples, such as anti-de Sitter space or
plane waves, are not globally hyperbolic. Moreover, as mentioned in
the Introduction, assuming the spacetime is globally hyperbolic will
not in general prevent spacelike quotients from leading to CTCs. We
will therefore focus on the weaker condition of stable causality.

\subsec{Quotients and causal inheritance}

Generally, it would be interesting to know when a 
quotient of a spacetime satisfying one of these conditions 
continues to satisfy that condition. Clearly, the quotient of a causal
spacetime along an isometry which is timelike or null somewhere will
fail to be causal, as we are explicitly introducing closed timelike or
null curves. However, as the example in the previous section
illustrated, choosing an isometry which is everywhere spacelike is
only a necessary condition to preserve the causal condition, and not a
sufficient one. So we would like to ask what more we can do. 

For a given spacetime, we can always adopt the brute force
method. Take a given isometry and perform the quotient. In the
resulting geometry, analyze the causal curves to check explicitly
whether or not there are CTCs.  While this at least in principle
provides a general solution of the problem of determining which
quotients satisfy the causal condition, it does not provide us with
much intuition. It would be preferable to have a more geometric
criterion. At the same time it would be more useful to have a
condition for the preservation of stable causality, since we have
argued that this is the condition we should generically impose on
physical spacetimes. 

Fortunately, there is a simple criterion. {\it Let $M/G$ be the quotient of a
stably causal spacetime $M$ by a subgroup $G$ of its isometry group
generated by some Killing vector $\xi^a$. Then $M/G$ 
is itself stably causal iff there
exists a time function $\t$ on $M$ such that $\CL_\xi \, \t = 0$.} That is,
we require that the quotient only relates points lying in the same
level surface of some time function $\t$; this time function is
invariant under the isometry we wish to quotient along. 

{\it Proof:} \ First, imagine that $M$ admits such a time
function. Then $\t$ will be single-valued on the quotient $M/G$, and
hence defines a time function on $M/G$. Thus, $M/G$ is stably
causal. Now suppose $M/G$ is stably causal. Then take any time
function $\t$ on $M/G$. We can pull this back to a function on $M$,
which will again be a time function, and trivially satisfies the
condition $\CL_\xi \, \t = 0$. QED.

One attractive feature of this condition is that it has the same form
as the conditions for the preservation of other structures: for
example, a quotient will preserve the symmetry associated with some
other Killing vector $\zeta^a$ iff $\CL_\xi \, \zeta= [\xi,\zeta]=0$, 
and it will
preserve the supersymmetry associated with a Killing spinor $\epsilon$
iff the spinorial Lie derivative $\CL_\xi \, \epsilon = 0$ \fapap. It
is perhaps slightly surprising that we can formulate a local condition
expressing the invariance under an isometry of a causality condition,
since the causality condition expresses a global property of the
geometry, like the absence of CTCs. The beauty of the formulation of
stable causality in terms of time functions is that it hides all the
global aspects in the time function.

Note that we only require that there exist at least one time function
such that $\CL_\xi \, \t = 0$; this need not be true for any arbitrary time
function we happen to consider. This makes this condition difficult to
disprove, as there seems to be too much freedom in the choice of time
function on $M$ to allow us to check it systematically. Hence, in
practice we will only use this condition to show that stable causality
is in fact preserved in some examples; where stable causality fails to be
preserved, we will demonstrate the failure by the brute-force method.

\newsec{Time functions for plane waves}

The specific example to which we wish to apply these ideas is the
plane wave spacetime. We will now show that plane wave spacetimes are
stably causal by explicitly writing down time functions for them. This
shows that we can apply our criterion to their quotients; it will also
develop important intuition on the construction of time functions.

Plane wave spacetimes are characterised by a covariantly constant null
Killing vector, along with full planar symmetry in the transverse
directions. One can write the general metric as
\eqn\plane{
ds^2 = -2 \, du \, dv - A_{ij}(u) \, x^i x^j \, du^2 +  dx^i \, dx^i}
where $A_{ij}$ are arbitrary functions of $u$ (although the equations
of motion will impose some constraint on the trace $A_{ii}(u)$), and
$i,j=1, \ldots, d$. We will proceed by first giving a time function for
the BFHP plane wave, where there is an obvious choice. We will then
generalise this to the general plane wave.

\subsec{Time functions for the BFHP plane wave}

The BFHP plane wave is the ten-dimensional metric (so $d=8$) with
$A_{ij}(u) = \mu^2 \delta_{ij}$. For convenience, we perform the
coordinate transformation to set $\mu=1$; the metric is then 
\eqn\bfhpmetrep{
ds^2 = -2 \, du \, dv -  x^i x^i \,du^2 + dx^i \,dx^i .
}
The only obvious guess for a time function is the coordinate $u$,
which is indeed treated as the `light-cone time' in studies of string
theory on this background. However, since the gradient $\nabla_a u$ is
null, this does not provide a good time function. Since it is null,
one might guess that there will be good time functions which only
differ from $u$ by `a little bit'. We will see that this is in fact
true, but we will arrive at our time function by a somewhat different
route. We want to exploit the fact that the metric \bfhpmetrep\ is
conformally flat, and thus can be mapped to the ESU. The ESU is stably
causal, and has an obvious time function, namely the global time in the
usual coordinate system. We can pull this function back to obtain a
good time function on the entire BFHP plane wave spacetime.

By making a sequence of coordinate transformations as  in \bena, starting 
with
\eqn\coodtra{
U  = \tan u \ , \qquad V = -v - { x^i \, x^i  \, \tan u \over 2}\ ,  
\qquad X^i = {x^i  \over \cos u}  \ , 
}
followed by 
\eqn\coodtrb{\eqalign{
2 \, U &= \( \tan{\psi + \zeta \over 2} - \tan{ \psi - \zeta \over 2} \) 
\, \cos \theta + \tan{\psi + \zeta \over 2} + \tan{ \psi - \zeta \over 2} \cr 
4 \, V & = \(\tan{\psi + \zeta \over 2} - \tan{ \psi - \zeta \over 2} \) 
\, \cos \theta - \( \tan{\psi + \zeta \over 2} + \tan{ \psi - \zeta \over 2} \)
\cr 
2 \, \sqrt{X^i \, X^i} & = 
\(\tan{\psi + \zeta \over 2} - \tan{ \psi - \zeta \over 2}
\) \, \sin \theta \ , 
}}
and finally, 
\eqn\coodtrc{\eqalign{
\cos \zeta &= - \cos \alpha \, \cos \beta \cr
\cos \theta \, \sin \zeta & =  \cos \alpha \, \sin \beta \ , 
}}
we can rewrite the BFHP plane wave \bfhpmetrep\ in a form conducive to
comparison to the ESU:
\eqn\esubfhp{
ds^2 = {1 \over 4 \, |e^{i \psi} - \cos \alpha \,  e^{i\beta}|^2} \,
(-d\psi^2 + d\alpha^2 + \cos^2 \alpha \, d\beta^2 + \sin^2 \alpha \, 
d\Omega_7^2) \ .}
The part of the metric inside the parenthesis is the metric of the 
ten dimensional ESU, where we have decomposed the metric of the $\S^9$
into an $\S^7$ and the $\S^1$ parameterised by $\beta$. The conformal factor
${1 \over 4 \, |e^{i \psi} - \cos \alpha \,  e^{i\beta}|^2}$ blows up 
when $\alpha =0$ and $\psi = \beta + 2 \pi n $ for some integer $n$ --- 
this corresponds to the 
one-dimensional conformal boundary of the plane wave. 

The global time $\psi$ is clearly a good time function for the
ESU. Our claim is that $\t = \psi$ is a good time function for the
geometry \bfhpmetrep. To see this, note that 
\eqn\esubfhpinv{
g^{\psi \psi} = - 4 |e^{i\psi} - \cos \alpha \, e^{i\beta}|^2 \ ,}
so the one-form $d\psi$ is timelike everywhere, except where the
conformal factor diverges \ie, at the conformal boundary. Since the
conformal boundary is not part of the spacetime manifold, the function
$\psi$ provides the good time function that we sought. This time
function can be written in terms of the coordinate chart used in
\bfhpmetrep\ as
\eqn\psinontriv{\eqalign{
\t(u,v,x^i) &= \tan^{-1} \left[ {(1+x^i x^i) \over 2} \, \tan u \, + \, v \,
+ \sqrt{ \left[ { ( 1- x^i x^i) \over 2} \, \tan u \, - \, v \right]^2 + 
{x^i x^i \over \cos^2 u}  } \; \right]  \cr
& \ \
+ \tan^{-1} \left[ {(1+x^i x^i) \over 2} \, \tan u \, + \, v \, -
\sqrt{ \left[ {( 1- x^i x^i) \over 2} \, \tan u \, - \, v \right]^2 + 
{x^i x^i \over \cos^2 u}  } \; \right] \cr
& = u + \tan^{-1} \left( {2 \, v \over 1+ x^i x^i} \right) \ .
}}
We can also check that this is a time function directly in the Brinkmann
coordinates used in \bfhpmetrep: a short calculation gives
\eqn\normpsi{
\nabla_a\t(u,v,x^i) \, \nabla^a \t(u,v,x^i) = - {4 \over (1 + x^i x^i)^2 + 
4 \, v ^2} \ , }
which implies that $\nabla_a \t$ is timelike for all finite values of
the coordinates. $\t(u,v,x^i)$ is also a continuous function of its
arguments, so it is a global time function\foot{In passing, we note
that this calculation also shows that the conformal factor relating
the plane wave to the ESU can be rewritten as ${4 \over (1 + x^i
x^i)^2 + 4 \, v^2}$.}.

Since the second term in \psinontriv\ is a $\tan^{-1}$, we see that
this time function is indeed of the form $\t = u +$ `a little bit'; in
particular, the level surfaces of $\t$ approach the surfaces $u =
$ constant as $v$ or the $x^i$ go to infinity. Note also that this time
function satisfies $\partial_u \, \t(u,v,x^i) =1$; that is, relative
to the timelike Killing vector $\partial_u$, this time function
satisfies the condition we argued in the previous section could be
used to pick out some `natural' time functions for us. It is possible
to find other time functions by a simple deformation of
\psinontriv. Consider functions of the form $\tau(u,v,x^i) = u +
\tilde{\tau}(v,x^i)$.  One would expect that any function
$\tilde{\tau}(y)$ with $y = {2 v \over 1 + x^i x^i}$ that ``looks
like'' $\tan^{-1} y$, such as $\tanh y$ or $y \over 1 + |y|$, ought to
work; and indeed, explicit checks confirm this. For instance, $ \tau =
u + {a \, v \over 1 + x^i x^i + b \, |v|} $ is also a good time
function provided $0 < a \le 2$ and $b^2 \ge 2 a$ with $b > 0$.

\subsec{Time functions for general plane waves}

Having constructed an explicit time function for the BFHP plane wave 
\bfhpmetrep, we can now try to generalise this 
construction to other plane waves. In general, it is still true that
$u$ has a null gradient, so it is reasonable to look for a time
function by using the ansatz $\tau(u,v, x^i) = u +
\tilde{\tau}(v,x^i)$. For a general plane wave, we have the
metric \plane\ where $A_{ij}(u)$ is an arbitrary function of $u$.  For
functions $A_{ij}(u)$ that are bounded above, it is very simple to
generalise the time function \psinontriv\ found for the BFHP plane
wave. Consider for example the candidate time function
\eqn\gentfn{
\t(u,v,x^i) = u + {1 \over \alpha} \, \tan^{-1} \left( {2 \, \alpha \, v 
\over 1 + B_{ij}\, x^i \, x^j} \right) \ , }
where $B_{ij}$ is a constant matrix.  In order for $ \nabla_a \t$ to
be timelike, \ie, $\nabla_a \t \, \nabla^a \t < 0$, 
it suffices to pick $B_{ij}$ such that  
\eqn\tconds{\eqalign{
(A_{ij}(u) - B_{ij})\,  x^i \, x^j & \leq 0, \cr
(B_{ik}\,  B_{kj} - \alpha^2 \, B_{ij}) \, x^i\,  x^j  & \leq 0 \ .
}}
\ie, $B_{ij} - A_{ij}(u)$ and $B_{ij} - {B_{ik} B_{kj} \over \alpha^2}$ 
are positive definite metrics on the transverse $\R^d$. Now for any
{\it bounded above} functions $A_{ij}(u)$, we can always find a matrix
$B_{ij}$ and a constant $\alpha$ satisfying this requirement.

There is a simple physical reason why it is easy to find a time
function for plane waves with $A_{ij}(u)$ bounded above: we can 
bound the quadratic form $A_{ij}(u) \, x^i \, x^j$ by $\mu^2 \,
\delta_{ij}\, x^i \, x^j$ for some $\mu^2$. That is, there is some
maximally symmetric plane wave which provides a fiducial metric for
our physical plane wave; if $A_{ij}(u) \, x^i \, x^j$ is bounded by
$\mu^2 \,\delta_{ij}\, x^i \, x^j$, the light cones of the metric
\plane\ will be bounded by the light cones of this fiducial plane
wave, and the time function constructed previously for this maximally
symmetric plane wave gives a time function on the general plane wave.

The case where $A_{ij}(u)$ can diverge as a function of $u$ would
appear to be more difficult, and indeed in this case we will have to
give up on the ansatz $\tau(u,v, x^i) = u + \tilde{\tau}(v,x^i)$ used
in the previous case, and allow some more general dependence on
$u$. Let us isolate the divergence by writing $A_{ij}(u)\, x^i \, x^j
= f(u) \bar{A}_{ij}(u) \, x^i \, x^j$, where $f(u)$ carries any
divergence to $+\infty$ and $\bar{A}_{ij}(u)$ is bounded above as a
function of $u$. We will assume without loss of generality that $f(u)
\geq 0 \; \forall \, u$. One might expect that a sufficiently severe
divergence in the metric will be somehow reflected in the time
function. This leads us to adopt an ansatz which satisfies $
\partial_u \, \t = 1 + f(u)$. 

Having isolated the divergent piece, we assume the remaining part of
the time function behaves as before, adopting the following ansatz:
\eqn\udeptfn{
\t(u,v,x^i) = u + \int^u \, d \bar{u} \, f(\bar u) + {1 \over \alpha} \, 
\tan^{-1} \left( {2\,  \alpha \, v \over 1 + B_{ij} \, x^i \, x^j  }
\right) \ .
}
For such a time function, 
\eqn\utfnone{
\nabla_a \t = \( 1 + f(u) \) \, (du)_a + {2 \, \( 1+B_{ij} \, x^i \,
x^j \) \over  
\( 1+ B_{ij} \, x^i \, x^j\)^2 + 
4\, \alpha^2\, v^2} (dv)_a - {4 \, v \,   B_{ij} \, x^j
\over \( 1+B_{ij} \, x^i \, x^j
\)^2 + 4\, \alpha^2 \, v^2} \, (dx^i)_a \ .  }
Evaluating the norm of $\nabla_a \t $ we obtain
\eqn\normudep{\eqalign{
\nabla_a \t \, \nabla^a \t  
& = -{ 4\, \( 1+f(u) \) \over \(1+\, B_{ij} \, x^i \, x^j \)^2 + 4 \,
\alpha^2 \, v^2} \cr & \;\;\; - {4\, \( 1+B_{ij} \, x^i \, x^j \)^2
\over \[ \( 1+B_{ij} \, x^i \, x^j \)^2 + 4\, \alpha^2 \, v^2 \]^2}
\, \(B_{ij} \, x^i \, x^j\, \( 1+f(u) \) - f(u) \, \bar{A}(u,x^i)\) \cr 
& \;\;\; - {16 \, v^2
\over \[ \( 1+B_{ij} \, x^i \, x^j \)^2 + 4\, \alpha^2 \, v^2 \]^2} 
\( 4 \, \alpha^2 \,  \( 1+f(u) \) 
B_{ij} \, x^i \, x^j -  B_{ik} B_{kj} x^i x^j  \) \ .
}}
We can therefore make $\nabla_a \t$ timelike everywhere by satisfying
\eqn\genuconds{\eqalign{
\( B_{ij}  - \bar{A}_{ij}(u) \) \, x^i \, x^j  & \geq 0 , \cr 
\( \alpha^2 \, B_{ij} - B_{ik} \, B_{kj} \) \, x^i \, x^j  & \geq 0 ,
}}
as in \tconds. As before, it is easy to find $B_{ij}$ and $\alpha$
satisfying the above requirements so long as $\bar{A}_{ij}(u)$ is
bounded above. Thus, we have obtained time functions for all plane
waves, explicitly demonstrating that these spacetimes are stably
causal\foot{It was shown in \penr\ that plane waves are, on the other
hand, not globally hyperbolic. We remark in passing that Penrose's
argument only applies to the case where $A_{ij}$ has at least one
positive eigenvalue. The case where $A_{ij}$ is negative semidefinite
is globally hyperbolic---however, the nontrivial plane waves in this
class violate the energy conditions.}.

Note that if the divergence in $A_{ij}(u)$ is slower than ${1 \over
u-u_0}$, $\t$ will remain bounded. In this case it seems to make sense
to continue the time function through the singularity at $ u =
u_0$. On the other hand, if $A_{ij}(u)$ diverges faster than ${1 \over
u-u_0}$, $\t$ will diverge as $ u \to u_0$, and we only have a good
time function on the whole spacetime if we treat the singularity as
the `end of the universe' and don't try to evolve through it. The
singularity is weak in the sense of Tipler \elsch,\tip\ if $A_{ij}(u)$
diverges slower than $(u-u_0)^{-2}$, so it might appear that the time
functions we constructed are not good enough to account for all weak
singularities. It is possible that there might exist better time
functions for these cases. However, this weak singularity is unstable
in the sense that a generic stress tensor will diverge as one
approaches it unless $A_{ij}(u)$ diverges slower than $(u-u_0)^{-1}$
\justin.  Hence we have constructed time functions which cover what we
regard as the full physical spacetime region in all cases.

\newsec{Causal properties of pp-waves}

All plane waves are hence stably causal. What about pp-waves?
Consider the general pp-wave metric,
\eqn\ppwave{
ds^2 = -2 \, du \, dv - F(u,x^i) \, du^2 + dx^i \, dx^i \ .}
The key new possibility that allowing a general function $F(u,x^i)$
permits, compared to the previous plane wave case, is that this
function can diverge to $+\infty$ as a function of the $x^i$ faster
than a quadratic at large $x^i$, or at some finite $x^i$.  This
additional `opening out' of the light cones can make the metric less
well-behaved causally than the plane waves.

Spacetimes of the form \ppwave\ will be stably causal {\it iff} the
behaviour of the function $F(u,x^i)$ is ``subquadratic'' as a function
of $x^i$ \flsa. Subquadratic means $F(u,x^i) \le A_{ij}(u) \, x^i \, x^j
\ \forall \ x^i$ for some $A_{ij}(u)$.  If this is true, we see that the
light cones of \ppwave\ are bounded by the light cones of some
fiducial plane wave, and hence it is clear that there is a good time
function on the pp-wave. This is thus a sufficient condition for
stable causality; remarkably, it is also necessary. As was shown in
\flsa, if the pp-wave is not subquadratic, it will remain causal but
fail to be even weakly distinguishing.

We will give a simple general proof of this statement. If the pp-wave
is not subquadratic, either $F(u,x^i)$ diverges to $+\infty$ faster
than a quadratic term $x^i x^i$ in some direction at large $x^i$, or
$F(u,x^i)$ diverges to $+\infty$ at some finite $x^i$, which for
convenience, we take to be the origin. In \flsa\ it was shown that the
spacetimes of the form \ppwave\ are not distinguishing if $F(u,x^i)$
grows faster than $x^i x^i$ at large $x^i$. We will review this
argument and show that it can easily be extended to address
divergences at finite $x^i$.

The essential point in both cases is that for pp-waves of this kind,
the future of any point $(u_0,v_0,x^i_0)$ will be the entire region
$u>u_0$. In particular, this implies that any two points in the plane
$u = u_0$ have the same future. By a similar argument, they will have
the same pasts. Hence, the spacetime is not weakly distinguishing. To
demonstrate that the future of any point $(u_0,v_0,x^i_0)$ is the
entire region $u>u_0$, we need to find a timelike curve $\gamma$
connecting $(u_0,v_0,x^i_0)$ to $(u_1,v_1,x^i_1)$ for any $u_1>u_0$
and arbitrary $v_1,x^i_1$. We will show this is possible for infinitesimal
separation, $u_1 - u_0=\epsilon$. It then follows immediately that it
is true for finite separation.

For the case where $F(u,x^i)$ grows faster than $x^i x^i$ at large
$x^i$, we will use a slight simplification of the argument in
\flsa. We consider a path composed of three pieces; first, we go in a
straight line from $(u_0,v_0,x^i_0)$ to a point $(u_0 + \epsilon/4, V,
X^i)$. We then travel in a straight line from $(u_0 + \epsilon/4, V,
X^i)$ to $(u_0 + 3 \epsilon/4, -V', X^i)$. Finally, we travel in a
straight line from $(u_0 + 3 \epsilon/4, -V', X^i)$ to $(u_0 +
\epsilon , v_1, x^i_1)$. We assume that $V, V', |X|^2$ are large
positive numbers; in particular, assume $V, V' \gg |v_0|, |v_1|$, and
$|X| \gg |x_0|, |x_1|$. Then the conditions for these three segments
to be timelike reduce to
\eqn\conda{-{1 \over 2} \, \epsilon  \, V + |X|^2 < 0,}
\eqn\condb{\epsilon \, (V + V') - {1 \over 4} \, F(u_0, X^i) \, 
\epsilon^2 < 0,}
\eqn\condc{-{1 \over 2} \, \epsilon \, V' + |X|^2 < 0.}
We can use the first and third conditions to eliminate $V$ and $V'$ in
the second condition, which then relates $\epsilon$ and $X^i$:
\eqn\fincond{F(u_0, X^i) \, \epsilon^2 > 16 \, |X|^2.} 
If $F(u_0,x^i)$ grows faster than $|x|^2$ at large $x^i$, this
condition can be satisfied for arbitrary $\epsilon$ simply by choosing
$X^i$ large enough. This shows that the indicated path is timelike in
this case\foot{If $F(u, x^i)$ is quadratic at large $x^i$ we see that
this condition can still be satisfied, but it now requires a finite
separation $\epsilon = u_1 - u_0$, in agreement with previous
discussions of light cones for plane waves \maro.}.

This argument can be trivially extended to the case where $F(u_0,x^i)$
is singular at some finite position, which we take W.L.O.G.\ to be at
$x^i=0$. We simply take $X^i$ in the above argument to be a value near
the origin, rather than a large value. This modifies \conda-\condc\ to
read
\eqn\nconda{-{1 \over 2} \, \epsilon \, V + |x_0|^2 < 0,}
\eqn\ncondb{\epsilon \, (V + V') - {1 \over 4}\, F(u_0, X^i)\, \epsilon^2 < 0,}
\eqn\ncondc{- {1 \over 2} \, \epsilon \, V' + |x_1|^2 < 0.}
Again, eliminating $V$ and $V'$, the main condition we need to satisfy
is 
\eqn\nfincond{F(u_0, X^i) \, \epsilon^2 > 8 \, (|x_0|^2+ |x_1|^2),} 
which we can clearly satisfy for arbitrary $\epsilon$ by choosing
$X^i$ sufficiently close to the origin, so that $F(u_0, X^i)$ is big
enough. Thus, for all ``superquadratic'' pp-waves, we can construct a
timelike path linking any two points $(u_0,v_0,x^i_0)$ and
$(u_2,v_1,x^i_1)$ with $u_1 > u_0$. As a result, the future
$I^+(u_0,v_0,x^i_0) = \{ (u,v,x^i): u>u_0 \}$ for any
$v_0,x^i_0$. This completes the demonstration that all
``superquadratic'' pp-waves are not weakly distinguishing.

In the other case, where $F(u,x^i)$ is a subquadratic function of
$x^i$, we have already argued that the light cones of the pp-wave are
bounded by the light cones of a fiducial plane wave, so it must be
stably causal. It is straightforward to construct an explicit time
function in this case, as we will do now. As in the plane wave
discussion, let us isolate from $F(u,x^i)$ any divergence to
$+\infty$ as a function of $u$, writing $F(u,x^i) = f(u) \,
\bar{F}(u,x^i)$, where $\bar{F}(u,x^i)$ is a bounded function of $u$.

A suitable ansatz for a time function is then
\eqn\subqtfn{
\t =  u + \int^u \, d\bar{u} \, f(\bar u) + {1 \over \alpha} \, 
\tan^{-1} \left( {2 \alpha v \over 1 + G(x^i) }\right) \ .
}
It is easy to check that is a good time function for the geometry \ppwave\ 
so long as 
\eqn\subqconds{\eqalign{
G(x^i) - \bar{F}(u,x^i) & \geq 0  \ , \cr 
4\alpha^2 G(x^i) - \partial_i G(x^i) \, \partial_i G(x^i) & \geq 0 \ .
}}
Given that $\bar F(u, x^i)$ is bounded above as a function of $u$, and
a subquadratic function of $x^i$, we can satisfy the first condition
by choosing some suitably large, smooth function $G(x^i)$, which grows
at most as $|x|^2$ at large $x^i$. We will assume that $G(x^i)$ is
everywhere positive; we can then satisfy the second condition by
choosing $\alpha$ large enough (here the fact that $G$ does not grow
faster than $|x|^2$ is crucial).

We have thus seen that pp-waves are either stably causal or
nondistinguishing. Stable causality does not rule out singularities in
$F(u,x^i)$; it only requires that any divergence in $F(u,x^i)$ is to
$-\infty$. Curiously enough, $F(x^i) \to - \infty$ guarantees a
genuine curvature singularity, while $F(x^i) \to + \infty$ need not in
fact correspond to a singular spacetime \cstruct.  For example, the
stably causal pp-wave with $F(u,x^i) = -1/|x|^{d-2}$ is singular
(recall $d$ is the number of $x^i$). On the other hand, the opposite
sign, $F(u,x^i) = 1/|x|^{d-2}$, which was shown to be geodesically
complete in \cstruct, is nondistinguishing.  Similarly, the pp-wave
spacetime \malmaoz,
\eqn\mamo{
ds^2 = -2 \, du \, dv - \( \cosh x - \cos y \)\, du^2 + dx^2 + dy^2 +
dz^i\, dz^i \ , }
leads to a geodesically complete but nondistinguishing spacetime.
Note also that the fact that certain pp-waves are nondistinguishing
implies that the technique for constructing causal completions
described in \causc\ cannot be applied to them. In particular, the
conclusions of \cstruct, wherein it was claimed that the spacetime
\mamo\ has a one-dimensional causal boundary are thereby rendered
invalid; since \mamo\ is not distinguishing, it does not make sense to
talk about its causal boundary.

\newsec{Quotients of the BFHP plane wave}

We have demonstrated that general plane waves are stably causal; we
would now like to ask what happens to the causal properties of these
plane waves when we quotient them by some spacelike isometry. We will
focus on the classification of the quotients for the BFHP plane wave;
this is the case of greatest interest, and will also provide the
richest structure, since it has the largest isometry group. We will
comment briefly on the lessons of this detailed analysis for
quotients of more general plane waves in the conclusions. 

\subsec{Classification of spacelike isometries}

The BFHP plane wave in ten dimensions has metric
\eqn\bfhpmet{ ds^2 = -2 \, du \, dv - \mu^2 \, x^i   x^i \, du^2 + 
dx^i \, dx^i,}
which is supported by a null five-form flux
\eqn\bfhpflux{F^{(5)} = \mu \, du \wedge (dx^1 \wedge dx^2 \wedge dx^3 
\wedge dx^4 + dx^5 \wedge dx^6 \wedge dx^7 \wedge dx^8).
}
The solution has 30 Killing vectors.  Some of these, such as
translations along $u$ and $v$ and rotations in the transverse space,
are manifest in the coordinate chart used to write the metric
\bfhpmet. There are also translation symmetries in the transverse
directions that are not manifest in these coordinates.  Note that
while the metric admits an $SO(8)$ rotation symmetry, some of this is
broken by the fluxes that are necessary to support the metric. If we
support the metric by 5-form Ramond-Ramond flux, as above, then we
break the $SO(8)$ down to $SO(4) \times SO(4)$. However, if we choose
to consider the 28 supercharge solution supported by 3-form flux as in
\bero,\michelsonb, then we have only a $U(4)$ rotation symmetry.  We will
write the isometry algebra for the maximally supersymmetric solution
above.

The Killing vectors for the geometry \bfhpmet\ are given as:
\eqn\killvec{\eqalign{
\ku &= - \partial_u \cr
\kv & = \partial_v \cr
\ki & = - \cu \; \partial_{x^i} + \mu \, \su \;  x^i \, 
\p_v \ ,   \qquad i = 1, \cdots , 8 \cr 
\kis & = - \mu \, \su \; \partial_{x^i} - \mu^2 \, \cu \; 
 x^i \, \p_v\ ,   \qquad i = 1, \cdots , 8 \cr
\krotij & = x^i \, \p_j - x^j \, \p_i \ ,  \qquad {\rm both}\
i,j = 1, \cdots, 4,
{\rm or\ both }\ i,j = 5, \cdots, 8 \ . 
}} 
The algebra generated by the Killing vectors is 
\eqn\kvalgebra{\eqalign{
[\kv, \ku] & = 0 \ , \;\;\; [\kv, \ki]  = 0 \ , \;\;\;
[\kv, \kis]  = 0\ , \;\;\; [\kv, \krotij]  = 0 \ , \cr
[\ku, \ki] & = \kis \ , \;\;\; [\ku, \kis] = -\mu^2 \, \ki \ , 
\;\;\; [\ku , \krotij] =0 \ , \cr
[\ki, \kjs] & = - \mu^2 \, \delta_{ij} \, \kv \ , \cr
[\krotij, \kk] & = \delta_{jk} \, \ki - \delta_{ik} \, \kj \ , \;\;\; 
[\krotij, \kks]  = \delta_{jk} \, \kis - \delta_{ik} \, \kjs \ \cr 
[\krotij, \krotkl] & = \delta_{jk} \, \krotil - \delta_{ik} \, \krotjl - 
\delta_{jl} \, \krotik  + \delta_{il} \, \krotjk \ .
}}

Let us now consider a general linear combination of the Killing vectors, 
\eqn\genlckv{
\zeta = a \, \ku + b \, \kv + c_i \, \ki + d_i \,\kis + \om_{ij} \, \krotij \ .
}
The norm of the Killing vector can be calculated from the metric \bfhpmet\ to 
be
\eqn\genkvnorm{\eqalign{
|\zeta|^2 &= - a^2 \, \mu^2 \,x^i  x^i  \, +  \, 2 \, a \, \[ b \, + \, \mu \, 
\su \,c_i \, x^i - \mu^2 \, \cu \, d_i \, x^i \] \cr
& \qquad +  \[ \cu \, c_i + \mu \, \su \, d_i +2 \, \om_{ij} \, x^j \]^2 \ . 
}}
The first line vanishes when $a =0$ and the second line being a
perfect square is non-negative. This motivates a natural distinction
between the situation when $a$ is vanishing and when $a$ is
non-vanishing.  As we will see, this choice is also motivated by the
distinct causal properties of these two cases. So we divide our
isometries into two classes; the \clg\ class where $a \neq 0$ and the
\clp\ class where $ a = 0$. 

\subsec{Quotients by the \clg\ class of isometries}

The \clg\ class of spacelike isometries includes the quotient involved
in the construction of the supersymmetric G\"odel-like spacetime from
T-duality of the plane wave \bghv. Thus, we expect to find that at
least some of these quotients contain CTCs. In fact, it turns out that
all the quotients in the \clg\ class contain CTCs. 

First, let us see that there are spacelike isometries in this
class. For $a \neq 0$, the first term in \genkvnorm\ is negative. To
achieve a spacelike norm, we must balance this negative term against a
positive contribution coming from the part involving the rotation
matrices $\omega_{ij}$. We must require $\[ - a^2 \mu^2 \, \delta_{ij}
+ 4 \, \om_{ik} \, \om_{jk} \] x^i \, x^j \ge 0$; this implies that
the rotation matrix $\om_{ij}$ must be of maximal rank. Also, to make the
norm non-vanishing at the origin of the transverse space $x^i = 0$, we
must require either $b \neq 0$ or both $c_i$ and $d_j$ non-vanishing
for some $i,j$. Once these conditions are satisfied, the Killing
vector will be everywhere spacelike. 

The argument for the presence of CTCs in these quotients is simple.
Consider a point on the $u =0$ plane, say $ p = (u_0 =0 , v_0,
x^i_0)$. The future of $p$ contains all points with $ u \ge {\pi \over
\mu}$ \maro.  The Killing vector has a $\ku$
component, which is just translation along $u$. When we quotient by
this isometry we identify some points on its orbits. In particular, we
will identify some point $ q = (u_1 \ge {\pi \over \,\mu}, v_1,
x^i_1)$ with the point $p$. But $q$ is in the future of $p$, so these
two points are connected by a future directed timelike curve, thus
giving us a CTC. So whenever we have a $\ku$ component to the isometry
we have a CTC. The example mentioned in Section 1.1 was a prototype of
this phenomenon in the simple setting of the four-dimensional
maximally symmetric plane wave.

We now see that the presence of CTCs in such quotients is intimately
connected to the special features of the light cones in plane wave
spacetimes---that the future of the origin contains {\it all} points
with $u \geq \pi/\mu$. This provides some understanding of why CTCs
can appear in quotients involving spacelike isometries in the plane
waves but not, for example, in flat space
\foot{If we regard this plane wave as the Penrose limit
of the supersymmetric AdS$_5 \times \S^5$ solution, the AdS$_5 \times
\S^5$ isometry which gives $\xi_{e_u}$ is $\partial_t - \partial_\psi$,
while $\xi_{e_v}$ is obtained from $(\partial_t + \partial_\psi)/R^2$,
where the Penrose limit is $R \to \infty$. Thus, the Killing vector
$\zeta$ on the plane wave corresponds to a Killing vector $\xi_{AdS} =
a(\partial_t - \partial_\psi) +\omega_{ij} \xi_{M_{ij}}$ in the Penrose
limit. When $a \neq 0$, the quotient along this Killing vector leads to
CTCs in the AdS$_5 \times \S^5$ space--and in particular in its
boundary--for precisely the same reasons as above. We thank Vijay
Balasubramanian for raising this issue.}.

\subsec{Quotients by the \clp\ class of isometries}

Now consider the case when we have $ a= 0$. This set of 
isometries has no $\ku$ component, and we will show that all the
quotients by spacelike Killing vectors in this class inherit the
property of stable causality from the parent spacetime. 

Let us first classify the physically distinct spacelike Killing
vectors with $a=0$. A general isometry of this form is
\eqn\kvclpa{
\zeta_P = c_i \, \ki + d_i \, \kis + \om_{ij} \, \krotij + b \, \kv \ , 
}
with norm
\eqn\kvclpnorm{
|\zeta_P|^2 = \[ \cu \, c_i + \mu \, \su \, d_i +2\, \om_{ij} \, x^j \]^2 \ . 
}
Now $\zeta_P$ is guaranteed to be non-timelike, but for it to be 
spacelike at the origin of transverse space, we need $ c_i$ or $d_i$ 
to be non-vanishing. Moreover, since the trigonometric functions have 
zeros at multiples of ${\pi \over 2 \,  \mu}$, we need to have at least 
one $c_i$ and one $d_i$ non-vanishing simultaneously.

Some of these isometries are related by conjugation by the isometry
group, and should not be counted as physically distinct. Since our
main interest is in the causal structure, we will consider conjugation
by the full $SO(8)$ symmetry of the metric; thus, we are treating as
equivalent isometries which only differ by their action on the
fluxes. Let us use this freedom to bring the isometry to a canonical
form. By an $SO(8)$ rotation we can set $c_i$ for $i \neq 1 $ to
zero. Since the $\ki$ and $\kis$ differ only by a translation in $u$
and scale, we may translate $u$ so as to set $d_1$ to zero. A further
rotation which leaves the $x^1$ direction fixed may be used to set
$d_i$ for $ i \neq 2$ to zero.  We can then use the freedom to choose
the overall scale to set $c_1 =1$, and finally set $b =0$ by
conjugation with respect to $\kones$. Hence, the canonical form for
the spacelike isometries in the \clp\ class is
\eqn\kvclpmin{
\zeta_P = \kone  + \a \, \ktwos + \om_{ij} \, \krotij \ , \qquad 
\a \neq 0} 
where we have  redefined $d_2/c_1 = \a$ for notational convenience,
and absorbed $c_1$ into $\om_{ij}$. 

Since the CTCs in the \clg\ class clearly had their origin in the
translation in $u$, one might guess that the \clp\ class of isometries
will lead to stably causal spacetimes. This is indeed true. An elegant
way to show this is to observe that the (string frame) geometry
obtained under Kaluza-Klein reduction to nine dimensions along the
general quotient in the \clp\ class is a pp-wave. This follows from
the basic property of the \clp\ class, that the Killing vector
\kvclpmin\ does not involve $\partial_u$. As a consequence,
$\partial_v$ remains a covariantly constant null vector in the
Kaluza-Klein reduced spacetime. 

In more detail, since the Killing vector \kvclpmin\ does not involve
$\partial_u$, one can make a coordinate transformation to bring this Killing
vector to the form $\zeta_P^a = \dza$ without redefining $u$. In fact,
the coordinate transformation can be taken to have the form 
\eqn\gcoodcha{\eqalign{
x^i & = b^i_a(U,z) X^a + c^i(U,z) \ , \cr
v & = V + f(U,z,X^a) \ , \cr
u & = U
}}
where $a = 2, \ldots, d$. In terms of these adapted coordinates, the
ten-dimensional metric can then be rewritten as 
\eqn\kkmet{\eqalign{
ds^2 & = -2\, dU \, dV - k(U,X^c) \, dU^2 + l_a(U,X^c) \, dU \, dX^a 
+ h_{ab} (U, X^c) \, dX^a dX^b \cr 
& + g(U,X^c) \, \( dz + j(U,X^c) \, dU + m_a(U,X^c) \, dX^c \)^2 \ . 
}}
The first line of this gives the string-frame metric in nine
dimensions obtained if we consider the Kaluza-Klein reduction along
$\zeta_P$. The essential point is that, because $u = U$ and $v = V +
\ldots$, the only non-zero $g_{\mu V}$ metric component in the
nine-dimensional metric is still $g_{UV} = -1$. This implies that in
this metric, the vector $\dVa$ is covariantly constant and null. Thus,
the metric is a pp-wave\foot{Since the dilaton depends on $g(U,X^a)$,
in the Einstein frame metric $\dVa$ is no longer covariantly constant,
but continues to be a null isometry.}. 

Now we saw in the previous section that pp-waves can be divided into
two classes: the subquadratic ones, which are stably causal, and the
superquadratic ones, which are causal but not distinguishing. Which of
these can arise under this Kaluza-Klein reduction? In any
superquadratic pp-wave, the function $F(u,x^i)$ must diverge to
$+\infty$ either at finite $x^i$ or faster than a quadratic at large
$x^i$. This implies that the curvature $R_{+i+j} = \partial_i
\partial_j F(u,x^k)$ must also grow unboundedly. Now we are
considering the quotient of the smooth, constant curvature BFHP plane
wave by an everywhere spacelike isometry, whose proper length is
bounded below for any given $\alpha, \om_{ij}$. This cannot lead to a
Kaluza-Klein reduced geometry with unbounded curvatures. Hence
Kaluza-Klein reduction along $\zeta_P$ must give a subquadratic
pp-wave.

The subquadratic pp-waves are all stably causal, and hence have global
time functions defined on them. We can therefore construct a time
function satisfying $\CL_{\zeta_P} \t =0$ on the BFHP plane wave for
any Killing vector $\zeta_P$ in the \clp\ class by constructing the
Kaluza-Klein reduction, identifying a time function on the resulting
nine-dimensional pp-wave geometry, and pulling it back to the
ten-dimensional geometry. This shows that all Killing vectors of the
form \kvclpmin\ satisfy the condition in section 2.2, and hence that
all quotients in the \clp\ class are stably causal.

This discussion gives a procedure for obtaining suitable time
functions, but it may seem somewhat abstract. To illustrate the
foregoing general discussion, we will now explicitly construct time
functions satisfying the condition of section 2.2 for a subclass of
the Killing vectors \kvclpmin. We will start by considering the
isometry $\zeta_P$ in \kvclpmin\ with $\om_{ij} =0$. We will
subsequently show that the time function $\t$ we obtain for this case
will continue to satisfy $\CL_{\zeta_P} \t = 0$ with $\om_{ij} \neq
0$ for $i,j \neq 1,2$, as the time functions will be rotationally
invariant in the other directions. We thus obtain explicit time
functions for a large subclass of the \clp\ class. Explicit time
functions could also be constructed for the remaining cases, but they
would be considerably more messy.

Let us consider the spacelike isometry 
\eqn\clpspl{
\zeta_P = \kone + \a \, \ktwos \ ,  
}
with $ \a \neq 0$. The case $\a = {1 \over \mu}$ was discussed in
\michelson.  For general non-zero $\a$ we wish to find the geometry
resulting from quotienting \bfhpmet\ by \clpspl. To this end it is
useful to find a coordinate chart such that $\zeta_P^a = \dza$. This
is achieved by the following change of coordinates:
\eqn\coordchn{\eqalign{
x^1 &= - \cu \, z - \a \, \mu \, \su \, {w \over \sqrt{g(u)}} \cr 
x^2 &= - \a \, \mu \, \su \, z + \cu \,  {w \over \sqrt{g(u)}} \cr 
-v & = {1 \over 2} \, (1 - \a^2 \, \mu^2) \, \mu \, \su \, \cu \,  z^2 
+ \a \, \mu^2 \, z \, {w \over \sqrt{g(u)}} - V  \ ,
}} 
where
\eqn\defg{
g(u) \equiv \cus + \a^2  \, \mu^2 \, \sus \ .
}
In terms of the new coordinates we can rewrite the metric \bfhpmet\ as 
\eqn\newmet{\eqalign{
ds^2 &= - 2 \, du \, dV - \mu^2 \, [ w^2 +
 \sum_{i \neq 1,2} \, x^i \, x^i ] 
\, du^2 +  dw^2 + \sum_{i \neq 1,2} \, dx^i \, dx^i \cr 
& +  g(u) \, 
\(dz +  {\a \, \mu^2  \over g(u)^{3/2} }\, w \, du  \)^2  .
}}

The metric \newmet\ can in the Kaluza-Klein sense be thought of as a
nine-dimensional metric along with a dilaton and gauge field. The nine
dimensional metric in string frame is just the first line of
\newmet. This is clearly a plane wave metric written in standard
Brinkmann coordinates\foot{Note that the dilaton here is just a
function of $u$ and hence the Einstein frame metric in nine dimensions
is also a plane wave. Also, since the nine-dimensional metric is a
plane wave, this quotient is generically preserving 16
supersymmetries; it can preserve additional supersymmetries at special
values of $\a$, as discussed in
\michelson.  Note that once we consider $\om_{ij} \neq 0$, we will
generically break all the supersymmetry.}; we can therefore use the
analysis of section 3.2 to determine an appropriate time function for
the nine-dimensional part of the geometry. Since $z$ is a spacelike
isometry and $\CL_z \t =0$, this will then immediately lift to a good
time function for the ten-dimensional geometry \newmet.

Explicitly, a suitable time function in the coordinates of \newmet\ is 
\eqn\qtfn{
\t(u, V, w, x^i)=  u + {1 \over \mu} \tan^{-1} \left( {2 \, \mu \, V \over  
1 + \mu^2 \, w^2 + \sum_{i \neq 1,2} \,\mu^2 \,  (x^i x^i)  } 
\right) \ . 
}
Expressing this time function in the original coordinate system of
\bfhpmet, we see that it takes the form 
\eqn\qtfnoldc{
\t(u, v, x^i)= u + {1 \over \mu} \tan^{-1} \left(  
{2 \, \mu \, [ v + A_1(u,x^1,x^2)]
\over  1 +  \, \sum_{i \neq 1,2} \mu^2 \, (x^i x^i) +   
A_2(u,x^1,x^2) } \right) \ , 
}
where
\eqn\tfnmess{\eqalign{
A_1(u,x^1,x^2) & = {\mu \over  2 \, g(u)^2 } \, 
\[ \cu \, x^1 + \a \, \mu \, \su \, x^2 \] \; \times \cr 
& \qquad \[ (g(u) + \a^2 \, \mu^2 ) \, \su \, x^1 - \a \, \mu
\, (g(u) + 1) \, \cu \, x^2 \]  \cr
A_2(u,x^1,x^2 ) & =  
 {\mu^2 \over g(u)} \, \( 
\a \, \mu\, \su \, x^1  - \cu  \, x^2 \)^2 \ .
}}

This is somewhat more complicated than the time functions considered
previously, but it is straightforward to check that it is a good time
function on the plane wave geometry, and furthermore that it satisfies
the condition $\CL_{\zeta_P} \t = 0$ for $\zeta_P$ given by \kvclpmin\
even once we allow $\om_{ij} \neq 0$ for $i,j \neq 1,2$. This thus
provides explicit time functions for a substantial subclass of the
Killing vectors \kvclpmin. 

We have argued that all quotients in the \clp\ class lead to
subquadratic pp-wave metrics on Kaluza-Klein reduction to nine
dimensions. Hence all quotients in this class are stably causal. We
will not discuss the construction of time functions for the more
general case, as the expressions become considerably more
complicated. 

\newsec{Discussion}

The main aim of this paper has been to explore the recent discovery
\herdeiroold, \herdeiro, \bghv, \bdpr,
that there are quotients of the plane wave spacetime by everywhere
spacelike isometries which lead to CTCs. First, we note that this
implies that requiring the isometry to be spacelike is a necessary
but, disappointingly, not a sufficient condition for the quotient to
inherit the causal property ({\it i.e,}\ the absence of CTCs) from the
parent spacetime. Our first result was to find a necessary and
sufficient condition to preserve the property of stable causality
under a quotient. The condition is simply that there exist a time
function $\t$ on the parent spacetime $M$ which is invariant under our
isometry: $\CL_\xi \, \t = 0$, where $\xi^a$ is the Killing vector we
wish to quotient along. Since stable causality implies causality, this
provides a sufficient condition for the absence of CTCs in the
quotient.

Our other main result was to classify the quotients of the BFHP plane
wave by everywhere spacelike isometries. We showed that the quotient
by a spacelike isometry will introduce CTCs if and only if the
isometry includes some translation in $u$. This is simply because any
two points which are separated in $u$ by more than $\pi/\mu$ are
timelike separated, so if we identify points with images that are
translated in $u$, eventually we must identify timelike separated
points. For the cases where the isometry does not involve translation
in $u$, we applied the general condition we derived to show that the
quotients in this class all preserve stable causality. 

We have seen that CTCs in spacelike quotients can appear in two
circumstances. They can appear in plane wave spacetimes, where their
occurrence is connected to the special features of the light cones in
these solutions. They can also appear when we consider a spacetime $M$
which is itself the quotient of some parent spacetime $N$ along a
spacelike isometry, $M = N/G$, so that $M$ has a spacelike isometry
with non-contractible $\S^1$ orbits.A similar construction is
possible for any spacetime with a noncontractible odd-dimensional sphere
factor. We gave the construction for BTZ,
which is an example of this type, in section 1.1, but there are
clearly many more examples in this class which might be discussed. It
is not clear if these are the only circumstances in which such CTCs
can occur; it would be interesting to understand the possibilities
more systematically. In particular, it is tempting to conjecture:
{\it For any globally hyperbolic space-time $M$ with trivial $\pi_n(M)$ 
for $n$ odd, any quotient $M/G$ by a subgroup $G$ of the isometry group 
generated by an everywhere spacelike Killing vector $\xi^a$ will be causal}.  
That is, we conjecture that
spacelikeness of the Killing vector is a sufficient condition for the
absence of CTCs so long as the spacetime does not have any
noncontractible cycles of odd dimension. An interesting direction
for further exploration of this family of ideas is to investigate
whether or not the null brane class of quotients of flat space are
stably causal. Note that the examples discussed in \herdeiroold, are 
not globally hyperbolic (as is the case with many p-brane geometries) 
and hence are not a counter-example to our conjecture.

It is interesting to compare our classification of quotients for the
BFHP plane wave to the classification for flat space \figsim\ and the
investigation of AdS \simon. Unlike in those cases, we have not found
any qualitatively new features in the general quotients; there is no
analog of the null brane quotient for the plane wave. This is simply
because the canonical form for the isometry \kvclpmin\ involves just
translation-like and rotation generators. The quotients which preserve
stable causality are simple generalisations of the quotients studied
in \michelson. In particular, as in \michelson, Kaluza-Klein reduction
along the isometries which do not involve rotations gives rise to plane
wave metrics in nine dimensions. More general quotients in this class
lead to subquadratic pp-waves. 

It should be easy to extend this analysis to quotients of more general
plane waves. It seems natural to conjecture that there will be
quotients along spacelike isometries which lead to CTCs only when the
$A_{ij}(u)$ are actually constants\foot{The matrix $A_{ij}$ must also
have at least one positive eigenvalue, so that the light cones have
the same feature as in the BFHP case \maro.}, so that translation in
$u$ is an isometry, and the isometry we quotient along involves a
translation in $u$. A similar discussion should also be possible for
the stably causal pp-waves where $F(u,x^i)$ is independent of $u$,
in cases where they have such spacelike isometries.

In passing, we also noted that, as first demonstrated in \flsa, there
are pp-waves which are not stably causal; in fact, which are not even
weakly distinguishing. This is interesting for two reasons: There have
been efforts to study string theory on such pp-waves
\malmaoz, \ruts, \baso. 
Although the analysis works in the usual way in light-cone
gauge, the fact that the spacetime is not weakly distinguishing should
encourage us to treat such analyses with some care. On the other hand,
it is interesting to have a natural example of a spacetime which fails
to be distinguishing; the textbook examples of such partial failures
of the causality conditions tend to be quite artificial. Perhaps
further study of these spacetimes will enable us to learn more about
the consequences of violating some of the causality conditions. 

\centerline{\bf Acknowledgements}

We would like to thank Liat Maoz and Jeremy Michelson for discussions.
We are grateful to the organisers of the workshop on String theory and 
Quantum Gravity in Amsterdam  
for their hospitality during the concluding stages of this project.
VH is supported by NSF Grant PHY-9870115, while MR acknowledges support 
from the Berkeley Center for Theoretical Physics and also partial support 
from the DOE grant DE-AC03-76SF00098 and the NSF grant
PHY-0098840. SFR is supported by an EPSRC Advanced Fellowship.


\listrefs
\end